\pdfoutput=1
\documentclass{iau}
\usepackage{graphicx}
\title[What we talk about when we talk about fields] 
{What we talk about when we talk\\ about fields}

\author[Ewan Cameron]   
{Ewan Cameron$^1$}

\affiliation{$^1$Department of Zoology, University of Oxford,
  Tinbergen Building, South Parks Road, Oxford, OX1 3PS, United
  Kingdom \\ email: {\tt dr.ewan.cameron@gmail.com}\\ website: {\tt astrostatistics.wordpress.com}}
\pubyear{2014}
\volume{306}  
\pagerange{xx--xx}
\jname{Statistical Challenges in 21st Century Cosmology}
\editors{Alan Heavens, Jean-Luc Starck \& Alberto Krone-Martins, eds.}

\begin{document}

\maketitle

\begin{abstract}
In astronomical and cosmological studies one often wishes to infer
some properties of an infinite-dimensional field indexed within a
finite-dimensional metric space given only a finite collection of noisy observational data.  Bayesian inference offers an
increasingly-popular strategy to overcome the inherent ill-posedness
of this signal reconstruction challenge.  However, there remains a great
deal of confusion within the astronomical community regarding the
appropriate mathematical devices for framing such analyses and the
diversity of available computational procedures for recovering posterior
functionals. In this brief research note I will attempt to clarify both
these issues from an ``applied statistics'' perpective,
with insights garnered from my post-astronomy experiences as a computational Bayesian
/ epidemiological geostatistician.

\keywords{Methods: data analysis--methods: statistical}
\end{abstract}

\firstsection 

\section{Introduction}
The potential afforded by Bayesian techniques for inferring the properties of
infinite-dimensional mathematical structures, such as 
random fields (to be understood here as random functions defined at each point of some
finite-dimensional metric space), has long been
recognised by both probability theorists, e.g.\ \cite{oha78}, and
practitioners: with the first wave of practical applications in
 geoscience (e.g.\ \cite{omr87}, \cite{han93}) and machine learning (e.g.\ \cite{ras96},
\cite{nea97}) contemporaneous with the advent of sufficiently
powerful desktop computers.  Cosmologists were at this time notable as \textit{`early
adopters'} and pioneers of the new techniques for field inference.
Indeed, the Monte Carlo methods for constrained simulation from
Gaussian random fields developed by \cite{ber87} and
\cite{hof91} remain key tools for efficient conditional
simulation, cf.\ \cite{dou10}.  

However, over the past decade the
sophistication of statistical analysis techniques brought to bear on
the study of 
cosmological fields has not kept pace with progress outside of
astronomy.  With modern tools such as the Integrated Nested Laplace
Approximation (INLA; \cite{rue09}), `variational inference'
(\cite{hen13}), particle filtering (\cite{del07}), and Approximate Bayesian
Computation (ABC; \cite{mar03}) almost entirely ignored to-date by the
cosmological community we have, in my opinion,
become the \textit{`laggards'} of the technology adoption lifecycle.

There are multiple factors seemingly to blame for this divergence: (i)
the
emergence of an 
isolationist attitude to the practice of cosmological
statistics; (ii) an over-emphasis on the
path integration-based conceptulisation of random fields, rather than
the measure-theory-based mathematics of mainstream statistics; and (iii) an
under-appreciation of the potential for stochastic process priors
(including, \textit{but not limited to}, the Gaussian process) as
flexible modelling components 
within the hierarchical Bayesian framework.  With the first already being well
fought back against by
inter-disciplinary programming in conferences such
as the IAUS306 and the SCMA series I will therefore focus in this
proceedings (as in my contributed talk) on the latter two.  In
particular, I aim to clarify a number of mathematical concepts crucial to a high-level
understanding of Bayesian inference over random fields and measures (Section \ref{mathematics}),
and then to highlight just a few of the exciting techniques to have
recently emerged in this area (Section \ref{techniques}).

\section{The Mathematics of Bayesian Field Inference}\label{mathematics}
Cosmologists and astronomers already
well-versed in the practice and theory of Bayesian
statistics in the finite-dimensional setting will typically have one
of two contrasting experiences upon first attempting
to extend these ideas to infinite-dimensional inference problems.  The
pragmatist will happily observe that the mechanics of computation are little
changed (e.g.\ the Gaussian random field at finite sample points is
distributed just as the familiar multivariate Gaussian), while the cautious
theorist will more likely be overwhelmed by a first acquaintance with
measure-theoretic probability (i.e., probability triples and the
algebra of sets).  But ideally one will have both experiences,
since each offers an equally important perspective, as I discuss in this
Section.

\subsection{Distributions over Infinite-Dimensional Space}
In formal statistics the core of probabilistic computation is framed
within the language of measure theory: the key object being the
\textit{`probability triple'} of (i) a sample space, $\Omega$, i.e., some non-empty
set; (ii) a $\sigma$-algebra, $\Sigma$, i.e., a collection of
subsets of $\Omega$ with $\emptyset,\Omega \in
\Sigma$, closed under the formation of complements, countable unions
and intersections; and (iii) a probability measure, $P$, i.e., a
countably additive set function
from $\Sigma$ to [0,1] for which $P(\emptyset) =0 $ and $P(\Omega) =
1$.  In this context Carath\'eodory's Extension Theorem provides the theorist with the machinary to build complex
probability triples and forge a rigorous notion of random variables as
measures on the pre-images of Borel $\sigma$-algebra sets of the real numbers; and from
this to the familiar mechanics of probability densities defined with respect to
the Lebesgue measure (e.g.\ the standard Normal with $f(x) =
\frac{1}{\sqrt{2 \pi}} \exp -\frac{x^2}{2} dx$).  Nevertheless, with
the Lebesgue 
measure behaving intuitively as a product measure in $\mathcal{R}^n$,
and with Lebesgue and Riemann integration interchangable in practice for all but a
few rare cases, the pragmatist can safely ignore these theoretical foundations in the
study of `real-world' problems in finite-dimensional settings.

In the context of probabilistic inference over \textit{fields}, however, one must
proceed with care as there exists no equivalent to the Lebesgue
measure to serve as a natural reference for defining densities in an
infinite-dimensional Banach space (e.g.\ the $L^p$ function spaces).
Hence, for Bayesian analysis in infinite-dimensional space we must
be deliberate in our choice of reference measure, which we encode
into the prior.  Typically we will do this indirectly by
assigning as prior the implicit measure (or `law') belonging to a given
\textit{stochastic process} (e.g.\ a Gaussian process, or Poisson
process) having sample paths within the field space under study.
Although quite technical the distinction between this formal
statistical approach and the path integration-based language of
cosmological papers, e.g.\ \cite{ens09,kit11}, is important if we are to connect with,
\textit{and thereby benefit from}, the
rich body of applied statistics literature on infinite-dimensional
inference.  Worth noting also is that the measure-theoretic equivalent
of the probability density is the `Radon-Nikodym (R-N) derivative', with a
trivial but illustrative example being that of the R-N derivative of posterior against prior
given by the likelhood function divided by the marginal likelihood, c.f.\ \cite{cot09}.

Finally, the measure theoretic definition of a stochastic process is a
collection of random variables \textit{indexed} by a set; here all
points of the physical metric space over which the field problem is to
be studied.  A key theoretical tool for the construction of stochastic
processes, which is greatly illustrative of their behaviour, is 
Kolmogorov's Extension Theorem.  The theorem gives conditions under
which a rule for assigning the distributions of the finite-dimensional
projections of an infinite-dimensional indexing set can be considered
sufficient to define a proper stochastic process.  The key condition
here is one of \textit{mutual consistency} of `coarser-binned'
finite-dimensional projected distributions
with respect to `finer-binned' ones; this idea was well-understood by
the MaxEnt pioneers, cf.\ \cite{ski98}: ``the prior must depend $\ldots$ on the pixel size $h$ in such
a way that subsidiary pixelisation is immaterial''.  

\subsection{Hierarchical Bayesian Models with Stochastic Process Priors}
As mentioned earlier for the pragmatist the mechanics of Bayesian
field analysis need differ little from
those of finite-dimensional inference; especially when we are able to
write our prior--likelihood pairing via a stochastic process embedded 
within a \textit{hierarchical Bayesian model}.  A typical hierarchical
model for Bayesian field inference is the following from \cite{get10}
in the context of epidemiological geostatistics,
\begin{eqnarray}
\nonumber N^{+}_i &\sim& \mathrm{Bin}(N_i,p(x))\\
\nonumber p(x) &=& g^{-1}(f(x))\\
\nonumber f|\phi &\sim& \mathrm{GP}(M_\phi,C_\phi).
\end{eqnarray}
We can read this model from the top down to understand the
generative process for the data: binomial sampling
from a population of $N_i$ at each site with underlying prevalence
(probability), $p(x)$, depending on the location of the site, $x$.
The prevalence field, $p(x)$, is in turn the
realisation of a Gaussian random field, GP, transformed to the range,
(0,1), by the \textit{link function}, $g^{-1}(\cdot)$.  A prior on the
parameters of the mean and/or correlation function of the GP would be
a natural extension of this particular model; while natural extensions
to other problems using the GP could include, e.g.,
exponentation of $f$ via $g^{-1}(\cdot)$, use of a Poisson process
likelihood function, or some penalisation of the GP sample paths. An
even greater diversity of hierarchical forms can then be built with
the addition of \textit{non}-Gaussian processes: cf.\ systematic error analysis (\cite{bur05})
and \textit{non}-proportional hazards modelling (\cite{ior09}) with Dirichlet
processes, online Bayesian classification with Mondrian processes 
(\cite{lak13}), and so on (Poisson processes, Gamma processes, Negative binomial
processes, etc.).

For complicated real-world problems
there is rarely an analytical solution to the resulting posterior 
so, as in ordinary Bayesian analysis, one will almost always turn to
computational methods.  Some of these are well-known to cosmologists
already, such as Gibbs sampling (e.g.\ \cite{wan04}) and Hamiltonian Monte Carlo (e.g.\ \cite{haj07,jas10});
however, a much greater number of specialised techniques for Bayesian field
inference remain largely undiscovered.  I highlight just a few of these briefly in the
following Section.

\section{Some Computational Techniques for Bayesian Field Inference}\label{techniques}
From a geostatistical perspective it is difficult to overstate the
revolutionary impact lately effected by the emergence of 
INLA (\cite{rue09}) and the stochastic partial differential equation (SPDE)
approach to Gaussian processes (\cite{lin11}).  Using the machinery
of the finite-element method already familiar to astronomers, the SPDE
approach aims to identify discretely-indexed Gaussian \textit{Markov} random fields
 providing weak approximations to their continuously-indexed
counterparts. The result is a system for approximate Bayesian
inference over random fields amenable to \textit{fast} computation via
sparse matrix operations; whereas Hamiltonian MCMC over field posteriors
is typically the reserve of cluster-computing, the INLA method enables
approximate field inference on desktop computers with run-times small
enough to allow for the important follow-up inference steps of model testing and
refinement.  A first glimpse of the potential for INLA in cosmological
applications was provided in a 2010 study of CMB reconstruction by
\cite{wil10}; but as yet this pioneering work remains unappreciated
(or, at least, uncited).

Another seminal technique for `Big Data' Bayesian inference over random
fields is that of `variational inference' (\cite{hen13}), in which
approximation to the full field posterior is made using sets of
data-driven `inducing points'.  Ongoing research efforts into the
application of Approximate Bayesian Computation (\cite{sou13}) and
Kalman filtering (\cite{sar14}) in the spatial domain are also well worth
keeping an eye on.


\begin{thebibliography}{}
\bibitem[Bertschinger (1987)]{ber87}
{Bertschinger, E.} 1987, 
\textit{ApJ}, 323, L103--L106

\bibitem[Burr \& Doss (2005)]{bur05}
{Burr, D., \& Doss, H.} 2005, 
\textit{J. Am. Statist. Assoc.}, 100, 242--251

\bibitem[Cotter et al. (2009)]{cot09}
{Cotter, S.L., Dashti, M., Robinson, J.C., \& Stuart, A.M.} 2009, 
\textit{Inverse Probl.}, 25, 115008

\bibitem[De Iorio et al. (2009)]{ior09}
{De Iorio, M., Johnson, W.O., M\"uller, P., \& Rosner, G.L.} 2009, 
\textit{Biometrics}, 65, 762--771

\bibitem[Del Moral et al. (2007)]{del07}
{Del Moral, P., Doucet, A., \& Jasra, A.} 2007, 
`Sequential Monte Carlo for Bayesian Computation' in 
\textit{Bayesian Statistics 8}, J.M. Bernardo, M.J. Bayarri,
J.O. Berger, A.P. Dawid, A.F.M. Smith \& M. West, eds., OUP, 1--34

\bibitem[Doucet (2010)]{dou10}
{Doucet, A.} 2010, `A Note on Efficient Conditional Simulation of
Gaussian Distributions',
Departments of Computer Science and Statistics, University of British Columbia

\bibitem[En\ss lin et al. (2009)]{ens09}
{En\ss lin, T.A., Frommert, M., \& Kitaura, F.S.} 2009, 
\textit{Phys. Rev. D}, 80, 105005

\bibitem[Handcock \& Stein (1993)]{han93}
{Handcock, M.S., \& Stein, M.L.} 1993, 
\textit{Technometrics}, 35, 403--410

\bibitem[Gething et al. (2010)]{get10}
{Gething, P.W., Patil, A.P., \& Hay, S.I.} 2010, 
\textit{PLOS Computat. Biol.}, 6, e1000724

\bibitem[Hajian (2007)]{haj07}
{Hajian, A.} 2007, 
\textit{Phys. Rev. D.}, 75, 083525

\bibitem[Hensman et al. (2013)]{hen13}
{Hensman, J., Fusi, N., \& Lawrence, N.D.} 2013, 
`Gaussian Processes for Big Data' in 
\textit{Association for Uncertainty in Artificial Intelligence}, UAI2013, 244

\bibitem[Hoffman \& Ribak (1991)]{hof91}
{Hoffman, Y., \& Ribak, E.} 1991, 
\textit{ApJ}, 380, L5--L8

\bibitem[Jasche \& Kitaura (2010)]{jas10}
{Jasche, J., \& Kitaura, F.S.} 2010, 
\textit{MNRAS}, 407, 29--42

\bibitem[Kitching \& Taylor (2011)]{kit11}
{Kitching, T.D., \& Taylor, A.N.} 2011, 
\textit{MNRAS}, 410, 1677--1686

\bibitem[Lakshminarayanan et al. (2013)]{lak13}
{Lakshminarayanan, B., Roy, D.M., \& Teh, Y.W.} 2013, 
`Top-down Particle Filtering for Bayesian Decision Trees' in 
\textit{International Conference on Machine Learning}, ICML2013

\bibitem[Lindgren et al. (2011)]{lin11}
{Lingren, F., Rue, H., \& Lindstr\"om, J.} 2011, 
\textit{J. R. Statist. Soc. B}, 73, 423--498

\bibitem[Marjoram et al. (2003)]{mar03}
{Marjoram, P., Molitor, J., Plagnol, V., \& Tavar\'e, S.} 2003, 
\textit{PNAS}, 100, 15324--15328

\bibitem[Neal (1997)]{nea97}
{Neal, R.M.} 1997, 
`Monte Carlo Implementation of Gaussian Process Models for Bayesian
Regression and Classification', 
\textit{Tech. Rep. 9702},
Department of Statistics, University of Toronto

\bibitem[O'Hagan (1978)]{oha78}
{O'Hagan, A.} 1978, 
\textit{J. R. Statist. Soc. B}, 40, 1--42

\bibitem[Omre (1987)]{omr87}
{Omre, H.} 1987, 
\textit{Math. Geol.}, 19, 25--39

\bibitem[Rasmussen \& Williams (1996)]{ras96}
{Rasmussen, C.E., \& Williams, C.K.I.} 1996, 
`Gaussian Processes for Regression' in
\textit{Advances in Neural Information Processing Systems 8},
eds. D.S. Touretzky, M.C. Mozer, M.E. Hasselmo, MIT Press, 514--520

\bibitem[Rue et al. (2009)]{rue09}
{Rue, Y., Martino, S., \& Chopin, N.} 2009, 
\textit{J. R. Statist. Soc. B}, 71, 319--392

\bibitem[S\"arkk\"a et al. (2014)]{sar14}
{S\"arkk\"a, S., Solin, A., \& Hartikainen, J.} 2014, 
`Spatio-Temporal Learning via Infinite-Dimensional Bayesian Filtering
and Smoothing', to appear in The IEEE Signal Processing Magazine

\bibitem[Skilling (1998)]{ski98}
{Skilling, J.} 1998, 
\textit{J. Microscop.}, 190, 28--36

\bibitem[Soubeyrand et al. (2013)]{sou13}
{Soubeyrand, S., Carpentier, F., Guiton, F., \& Klein, E.K.} 2013, 
\textit{Stat. Appl. Genet. Mol. Biol.}, 12, 17--37

\bibitem[Wandelt et al. (2004)]{wan04}
{Wandelt, B.D., Larson, D.L., \& Lakshminarayanan, A.} 2009, 
\textit{J. R. Statist. Soc. B}, 71, 319--392

\bibitem[Wilson \& Yoon (2010)]{wil10}
{Wilson, S.P., \& Yoon, J.} 2010, 
preprint(arXiv:1011.4018)

\end{thebibliography}
\end{document}